\newcommand{\CLASSINPUTtoptextmargin}{0.7505in}
\newcommand{\CLASSINPUTbottomtextmargin}{1.01in}
\documentclass[conference]{IEEEtran}
\IEEEoverridecommandlockouts

\usepackage{cite}
\usepackage{amsmath,amssymb,amsfonts}
\usepackage{algorithmic}
\usepackage{graphicx}
\usepackage{textcomp}
\usepackage{xcolor}
\usepackage{url}
\usepackage{subcaption}
\usepackage{cite}
\usepackage[font=small, labelfont=bf]{caption}
\usepackage{dblfloatfix}
\def\BibTeX{{\rm B\kern-.05em{\sc i\kern-.025em b}\kern-.08em
    T\kern-.1667em\lower.7ex\hbox{E}\kern-.125emX}}
\begin{document}
\title{Context-Aware Management of IoT Nodes: Balancing Informational Value with Energy Usage\\
}

\author{\IEEEauthorblockN{Nihal Ahmad$^{*,\dagger}$, Talha Manzoor$^{*,\dagger}$, Ijaz Haider Naqvi$^{\dagger}$}
\IEEEauthorblockA{\textit{$\dagger$ Department of Electrical Engineering, Lahore University of Management Sciences (LUMS), Lahore, Pakistan} \\
\textit{* Centre for Water Informatics and Technology, Lahore University of Management Sciences (LUMS), Lahore, Pakistan}\\
\{23060001, talha.manzoor, ijaznaqvi\}@lums.edu.pk}
}
\maketitle
\begin{abstract} 
The operational lifetime of energy-harvesting wireless sensor nodes is limited by availability of the energy source and the capacity of the installed energy buffer. When a sensor node depletes its energy reserves, manual intervention is often required to resume node operation. While lowering the duty cycle would help extend the network lifetime, this is often undesirable, especially in time-critical applications, where rapid collection and dissemination of information is vital. In this paper, we propose a context-aware energy management policy that helps balance the two opposing objectives of timely data collection and dissemination with energy conservation. We capture these objectives through the \textit{Value of Information (VoI)} of observations made by a sensor node and the \textit{State of Energy (SoE)} of the energy buffer. We formulate the energy management policy as a Model Predictive Control (MPC) problem which computes device sampling and transmission frequencies to maximize a defined utility criterion over a finite, receding, time-horizon. In the process, we also develop a unique mathematical representation for VoI, that adequately captures aspects related to continuity in monitoring, urgency of dissemination, and representation of the phenomena being observed. In the end, we use data collected from a real-world flash flood event, to evaluate our decision framework across multiple scenarios of energy availability.
\end{abstract}

\begin{IEEEkeywords}
Context-awareness, Energy Efficiency, Value of Information, Model Predictive Control, Adaptive Sampling.
\end{IEEEkeywords}

\section{Introduction}
Wireless sensor networks (WSNs) are ubiquitous in a host of applications \cite{ali2017comprehensive} due in large to their scalability and distributed nature. WSNs are extensively deployed in critical environmental applications where continuous monitoring and reporting of data is required. This corresponds to frequent sampling and transmission which contrasts with the objective of energy conservation for sensor nodes that already have limited energy storage to begin with. Energy harvesting nodes are able to offset this limitation to a large extent \cite{adu2018energy}. Depending on the application, energy may be harvested from multiple sources including solar radiation, vibrations of the mounting surface, tidal waves, or other phenomena associated with the process being observed. We concern ourselves with harvesting sources that are uncontrollable but predictable in nature (such as solar), i.e., they cannot be exploited to generate energy at any desired time but their behavior can be modeled fairly accurately to predict the energy harvest \cite{kansal2007power}. Due to the uncontrollable nature of the energy harvest, an effective energy management policy is required that simultaneously considers the contrasting objectives of continuous monitoring and energy conservation. 

Energy-efficient operation in IoT networks requires data to be prioritized according to significance to minimize energy expenditure on non-essential samples. Here, we adopt the notion of the Value of Information (VoI) as a metric to quantitatively rank sampled and transmitted data. It is important to note that no single, universally accepted definition of VoI exists in the literature. Rather, the quantitative approach involves assigning value to data attributes such as timeliness, content, format, and cost, depending on the application and context in which the sensor network is deployed \cite{ahituv1989assessing} \cite{bisdikian2013quality}. Effective VoI representations account for both information-theoretic dimensions (entropy, completeness, uncertainty, etc.) and application-oriented factors (urgency, impact, cost-effectiveness, etc.). Practical applications include flood monitoring \cite{kho2009decentralized}, glacial sensing \cite{padhy2006utility}, and vehicular networks \cite{giordani2019framework} (to name a few), each with their own formulations of the value functions which consider attributes ranging from forecasting ability and uncertainty reduction to age of data, and threat proximity. In this study, we couple VoI with energy conservation objectives for critical applications in remote locations. For such applications, prediction of future energy availability is necessary for optimizing node operations over a long-term horizon. Such approaches often rely on energy-source models (see representative studies \cite{moser2009adaptive,braten2019adaptive,donohoo2013context} and references found therein). However, effectively adapting in real-time to both the fluctuating energy source and the evolving demands of the sensing task remains a significant limitation.

The need for IoT devices that employ ``smart" energy optimization strategies has spawned a number of research efforts evident in areas such as adaptive sampling \cite{giouroukis2020survey}, context-aware computing \cite{perera2013context}, and self-adaptive IoT architectures \cite{alfonso2021self}. The contextual information being considered in the device management strategy defines the behavior of the device in practice. On one end of the spectrum lie approaches that focus entirely on the application or process-related context (e.g., FloodNet \cite{zhou2007floodnet} which uses a flood prediction algorithm to adjust its sampling rate) and are not concerned with energy conservation. On the other end there are approaches that concern themselves entirely with the energy harvesting profile and capacity of the energy storage buffer \cite{kansal2007power, vigorito2007adaptive}, but are not sensitive to the dynamic process-driven needs of the sensing application (for instance, observations during a flood event may be missed in favor of conserving energy). An approach worth mentioning here is a study \cite{zhang2015value} that formulates a multi-armed bandit problem with the goal of maximizing VoI under energy constraints. While the Kullback-Liebler divergence used to formulate VoI prioritizes unusual data with a surprise element, it does not directly incorporate other contextual aspects such as threat proximity or the ability to accurately reconstruct physical phenomena of interest.

\begin{figure}[t]
\vspace{0.02in}
\centering
\includegraphics[width=0.8\linewidth]{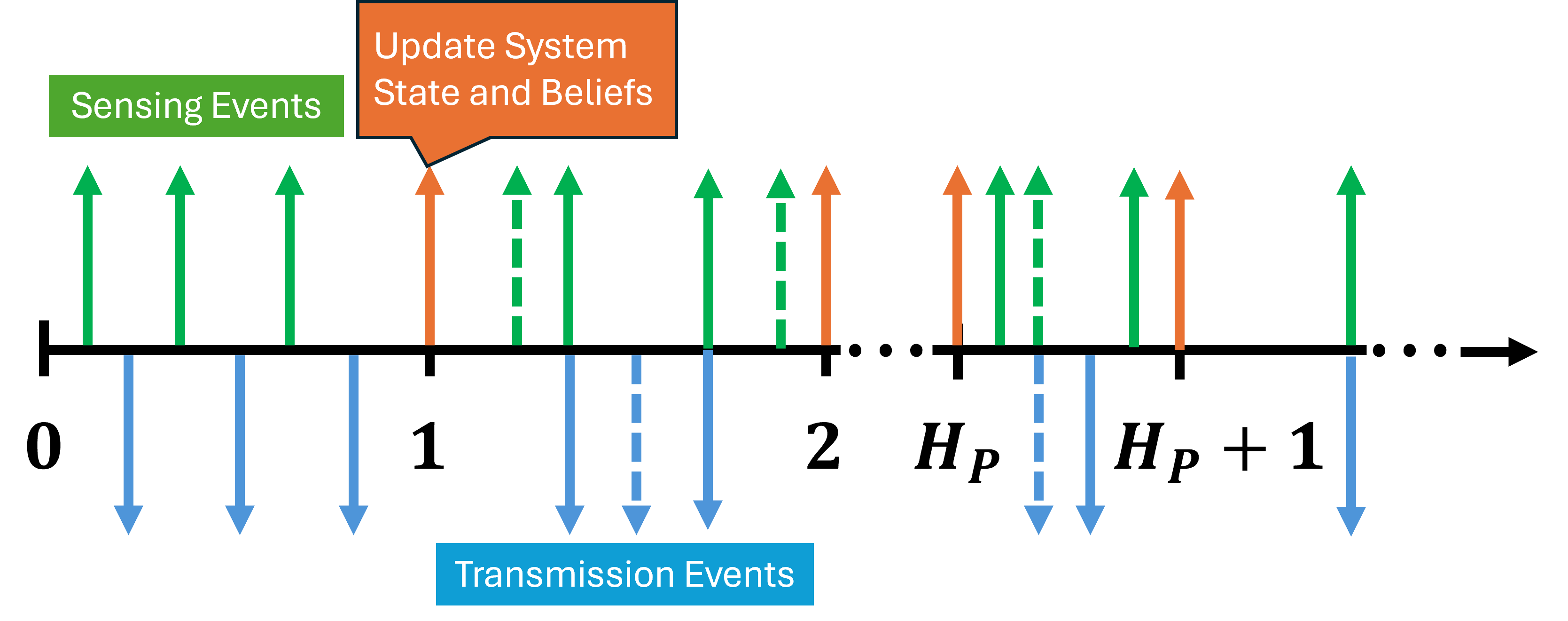}
\caption{Timing diagram with the prediction horizon $H_p$. The dashed arrows represent the old decisions before updating the beliefs while the solid arrows are the decisions made after updating beliefs and are actually followed.}
\label{fig:timing-diagram}
\vspace{-10pt}
\end{figure}

In this paper, we address the node management problem by modeling it as the strategic selection of appropriate sampling and data transmission frequencies. The frequency selection is formulated as a Model Predictive Control (MPC) problem \cite{kouvaritakis2016model}, to deal with inherent variability and uncertainty in the observed process and available energy. The MPC optimizes a specified notion of utility, combining a novel representation of VoI with the State of Energy (SoE) over a receding time horizon. We rely on exogenous specification of the predictions for the process being observed, and the future availability of energy for harvest. Since we do not focus on the predictive models for the process and energy availability, we take them to be piecewise continuous and updated at each decision cycle of the MPC. These reflect the beliefs of the human operator regarding the process and energy availability (which may come from a predictive model or human experience). Figure \ref{fig:timing-diagram} depicts a timing diagram that indicates the points at which the belief updates take place. 

In the following sections, we explain the decision framework in detail. We begin with the specification of the MPC optimization problem in Section \ref{sec:sys-model}. There, we also present the overall system model, our definition of utility, and its dependence on VoI and SoE. In Section \ref{sec:voi}, we derive our mathematical representation of VoI. We incorporate multiple process-related factors: 1) Threat rating, 2) Process fidelity, and 3) Cost of update delay. The representation is versatile across various applications and processes. We illustrate the VoI against data observed from a real-world sensor network under different preferences attached to the 3 VoI factors. Next in Section \ref{sec:soe} we describe our representation of the energy conservation objective via State of Energy (SoE), which directly corresponds to the longevity of device operation. We discuss the battery model, the hardware setup used for energy profiling of the sampling and transmission operations, and our method for SoE estimation. Finally, we evaluate the proposed decision framework under multiple scenarios for data gathered during a real-world flash flood event. This is presented in Section \ref{sec:results}. The results demonstrate that our framework dynamically adjusts the sampling and transmission frequencies in a manner that is responsive to both process and energy contexts, depending on the situation. We compare with fixed-frequency behavior at both high and low rates. We conclude in Section \ref{sec:conclusion}.
\section{System Model}\label{sec:sys-model}
\begin{figure}[t]
\centering
\vspace{0.08in}
\includegraphics[width=0.9\linewidth, height=5cm, keepaspectratio]{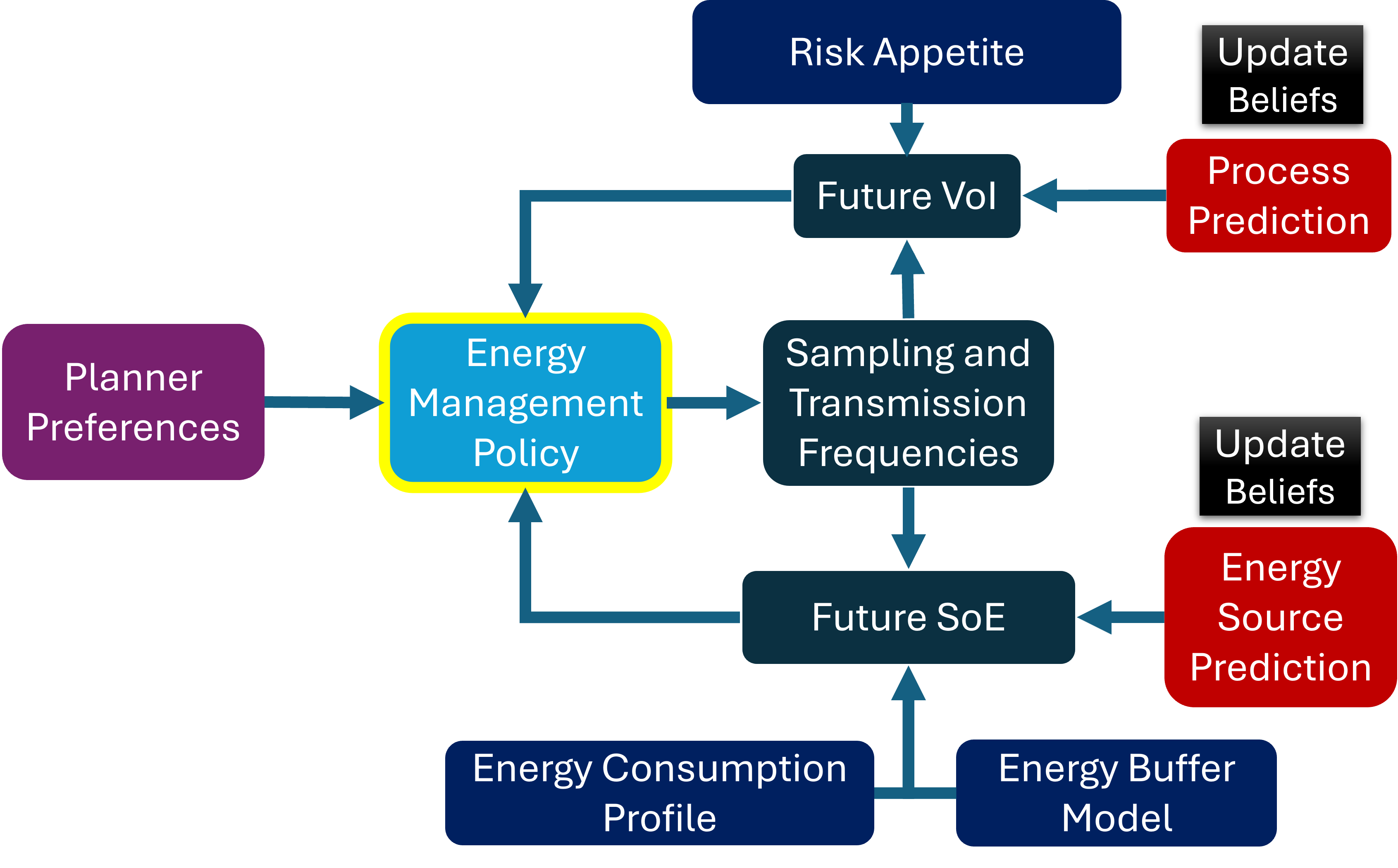}
\vspace{-0.1in}
\caption{Our system model depicting the flow of contextual information to the energy management policy. The selected frequencies in turn affect the projected State of Energy and Value of Information. }
\label{fig:system-model}
\vspace{-15pt}
\end{figure}

Consider a stationary wireless sensor network consisting of a number of spatially distributed sensor nodes placed in a remote environment. Each node is equipped with a host of sensors for environmental monitoring, a rechargeable energy buffer, an energy harvesting module, and bidirectional communication capabilities.
The sensor node is able to measure certain physical properties from the environment, depending on the sensor used, and can transmit data to an information sink. The sink can receive data packets from the sensor node and process them for event detection. Figure \ref{fig:system-model} shows the flow of information for the proposed system. It is important to note that we assume knowledge about future energy availability and process dynamics (modeled through beliefs), the energy consumption profile for each task of the node, and the capacity of the energy storage buffer.
\vspace{-0pt}
\subsection{MPC Formulation}
Let $K=\{0, 1, 2,  \ldots, H_p\}$ be the discrete set that represents the time steps for which decisions are made, where $H_p$ is the prediction horizon.
The sampling and transmission frequencies in the window $k\in K$ are given by $f_s(k)$ and $f_t(k)$ respectively. Let $V_i(f_s(k), f_t(k))$ and $S_e(f_s(k),f_t(k))$ be the Value of Information and State of Energy of the node respectively.
Our objective is to find a set of decisions, $f_s(k)$ and $f_t(k)$ for all $k \in K$, that maximize the discounted utility over the prediction horizon $H_p$. The control problem is set up as
\vspace{-10pt}
\begin{equation*}
\begin{aligned}
    &\max_{f_s(k), f_t(k) \geq0} \sum_{k=0}^{H_p} \frac{1}{\left(1+\zeta\right)^k}U(f_s(k), f_t(k)),  \\
    &\text{s.t. } \forall \,\, k\in K\\ 
    &S_e(f_s(k), f_t(k)) \geq 0; \quad \Delta -\Delta f_s(k)d_{sn}-\Delta f_t(k)d_{tx} \geq 0; \\
    & f_s(k) \geq f_t(k); \quad f_s(k) \leq f_s^{\text{max}}; \quad f_t(k) \leq f_t^{\text{max}}.
\end{aligned}
\end{equation*}
where $U(f_s(k), f_t(k))$ is the total utility gained in the window $k$, $d_{sn}$ and $d_{tx}$ are the time durations for sampling and transmission respectively, $f_s^{\text{max}}$ and $f_t^{\text{max}}$ are upper bounds imposed by hardware and physical limits, $\zeta \geq 0$ is a positive discount factor, and $\Delta$ is the time-duration of the decision window. The process is repeated after shifting the decision window by a single step i.e. the control horizon is equal to the prediction horizon.
We define the utility function as\vspace{-2pt}
\begin{equation*}
    U(f_s(k), f_t(k)) = w_i \hspace{.2em} V_i(f_s(k), f_t(k)) + w_e \hspace{.2em} S_e(f_s(k),f_t(k)),
\vspace{-2pt}\end{equation*}
where $w_i>0$ and $w_e>0$ are the weights that reflect the decision-maker's preferences regarding the relative importance of the VoI factor as compared to the SoE factor. The VoI and SoE are normalized over their respective ranges.
\vspace{-3pt}
\section{Value of Information}\label{sec:voi}
We define VoI based on the inherent properties of the collected data and its context-dependent relevance. We combine three different VoI factors, recognizing that the value of the network and its information is determined by its utility in achieving application objectives through the provision of accurate and timely data. These factors are described below.
\vspace{-3pt}
\subsection{Value of Information Factors}
We assume the following factors to contribute to VoI in a sensor network:
\subsubsection{Threat Rating} The threat rating of an observation, is defined as the likelihood, observed from the process, of a harmful event occurring.
We represent it as the difference of the current observation from a pre-defined critical threshold. An observation close to or exceeding this threshold implies an event occurrence, and thus carries more value. The contribution of this factor to the VoI of an observation $x$ is represented by $v_c(\cdot)$ and is given as\vspace{-1pt}
\begin{equation}\label{eq:threat-rating}
     v_c(x) = e^{-\lambda_c(x_{c}-x)}u(x_{c} - x) + u(x-x_{c})
\vspace{-1pt}\end{equation}
where $x_c$ is the critical threshold, $\lambda_c>0$ reflects the planner's attitude towards the threat posed, and $u(\cdot)$ is a unit step function. A planner that is indifferent to the threat will have a large value of $\lambda_c$ and vice versa. 
\subsubsection{Process Fidelity}
Process fidelity represents the accuracy with which the process can be reconstructed from the observations made. The need to incorporate process fidelity arises from the fact that each successive observation provides less and less new information about the process. 
Consider a dynamical process with a set of modes, sampled at a frequency to capture the dominant modes. Oversampling leads to new redundant, correlated data, without introducing any new information. This is related to the fact that a process only needs to be sampled at a minimum Nyquist frequency for accurate reconstruction. Process fidelity represents the need for a large enough sampling rate that can be used to accurately reconstruct the process while avoiding oversampling. 
We use an exponential function to express the marginal contributions of observations. The contribution of the factor process fidelity to Value of Information for a specific selected sampling frequency $f_s$ is represented by $v_r(.)$ and given as
\begin{equation}
    v_r(f_s(k)) = 1-e^{-\alpha_r \Delta f_s(k)},
\vspace{-1pt}\end{equation}
where $\alpha_r>0$ is a parameter based on how fast or slow-moving the process is. A slow moving process only requires a small frequency for reconstruction and will therefore have a larger value of $\alpha_r$.

\subsubsection{Cost of Update Delay}
The Cost of Update Delay (CoUD) \cite{kosta2020cost} is a metric that characterizes the loss in information due to high transmission inter-arrival times. A common objective of sensor networks is to keep the end user up-to-date with the freshest information. Any delays can affect the response time of the human actuating units. The contribution of this factor to the VoI against a frequency $f_t(k)$ is represented by $v_d(\cdot)$ and is given as\vspace{-1pt}
\begin{equation}
    v_d(f_t(k))= D_oe^{-\alpha_d f_t(k)},
\vspace{-1pt}\end{equation}
where $\alpha_d>0$ is the parameter that affect how quickly the cost decreases with the frequency (a lower value of $\alpha_d$ imposes more cost on the same frequency), and $D_o>0$ is the maximum cost imposed due to infrequent transmissions.
\subsection{Mathematical Expression for Value of Information}
We now define VoI as a combination of the three factors defined above as follows\vspace{-4pt}
\begin{equation*}
V_i(f_s(\!k\!),\!f_t(\!k\!)) \!=\! v_c(\!x\!)v_r(f_s(\!k\!)/v_c(\!x\!))- v_c(\!x\!) v_d(f_t(\!k\!)/ v_c(\!x\!))
\end{equation*}
The formulation $v_r(f_s(k)/v_c(x))$ and $v_d(f_t(k)/v_c(x))$ ensures that the VoI is not just dependent on the number of observations made but is also contextually weighted. A higher threat rating will ensure that a higher sampling and transmission frequency is needed to achieve the same utility and vice versa. Substituting the expressions for the individual VoI factors gives us the final expression for the VoI as 
\begin{equation*}
 V_i(f_s(k),\!f_t(k)) \!=\! v_c(x)(1\!-\!e^{\!\frac{-\alpha_r \Delta f_s(k)}{v_c(x)}}\!)- v_c(x)D_oe^{\!\frac{-\alpha_df_t(k)}{v_c(x)}}
\end{equation*}
We consider the first term to be ``Value of Information Update" (VoIU) \cite{kosta2020cost}. VoIU describes the importance of the incoming sensed information towards fulfilling the objectives of the network. New information indicates new changes in the process and thus is always non-negative. However, note that the VoI can be non-positive for certain low transmission frequencies. This implies that at such low transmission frequencies, the data is transmitted such infrequently that the network essentially fails to fulfill its objectives. VoI function is concave as can be trivially determined from its Hessian.

\begin{figure}[t]
\includegraphics[width=1\linewidth]{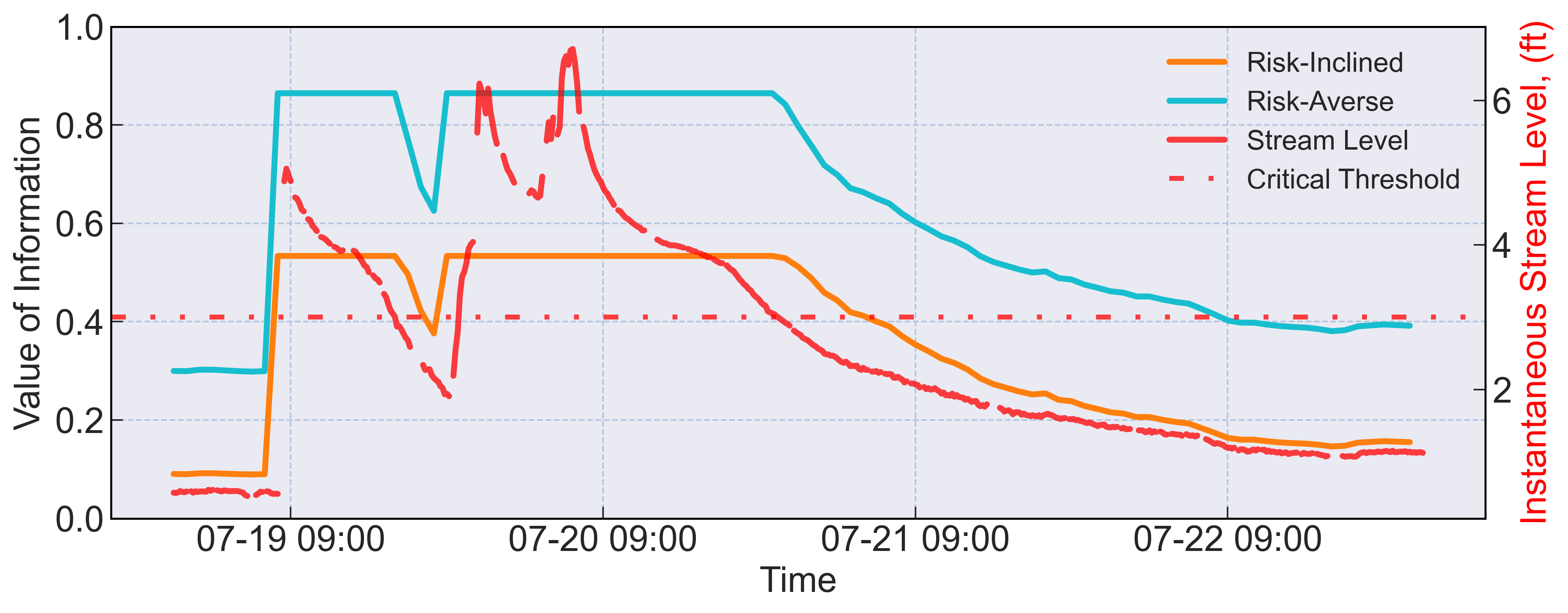}
\vspace{-20pt}
\caption{Behavior of a risk-inclined and a risk-averse planner with changing stream levels. The parameters for the risk-inclined planner are $\lambda_c = 1$, $\alpha_r=0.009$, $\alpha_d=0.025$, $D_o=0.5$ and for the risk-averse planner are $\lambda_c = 0.5$, $\alpha_r=0.02$, $\alpha_d=0.25$, $D_o=0.25$}
\label{fig:planner-behaviours}
\end{figure}
\vspace{-1pt}
\subsection{Value of Information and Risk Appetite}
The four parameters $\lambda_c$, $\alpha_r$, $\alpha_d$, and $D_o$ collectively represent the risk appetite of the planner. A risk-averse planner would have low values for $\lambda_c$, and $D_o$ and high values for $\alpha_r$ and $\alpha_d$. On the other hand, a risk-inclined planner would have high values for $\lambda_c$, and $D_o$ and low values for $\alpha_r$ and $\alpha_d$. Figure \ref{fig:planner-behaviours} compares the VoI for two planers with different risk appetites against data observed from a real-world stream gauge (details on the sensor network can be found elsewhere \cite{manzoor2022development}). The stream levels are measured at the Tarappi stream at an observation point in Lawa, Punjab, Pakistan, from 19th to 23rd July, 2021. We see that for both planners, the VoI increases as the threat proximity (the proximity of $x$ from $x_c$) increases. Moreover, the risk-averse planner has a consistently higher VoI than the risk-inclined planner. We expect this to result in higher sampling and transmission rates for the risk-averse planner to maintain a constant state of heightened alertness.

\vspace{-4pt}
\section{State of Energy}\label{sec:soe}
We represent the energy efficiency objective through the State of Energy (SoE). We treat SoE as a measure of how long a sensor node can operate before its stored energy is fully depleted. In the following text, we first define our battery model to represent the energy storage buffer. Next, we discuss the process for energy profiling for a real sensor node followed by the derivation of the SoE expression.
\vspace{-4pt}
\subsection{Battery Model}
Existing battery models can be classified into four main categories: electrochemical models, analytical models, electrical circuit models, and stochastic models.
A commonly used and simple analytical model is the Coulomb counting method. It estimates the State of Charge (SoC) by tracking the charge moving into and out of the battery. Let $0\leq z(k) \leq 1$ be the SoC at $k$, $C$ be the total charge capacity of the battery and $I(k)$ be the current drawn from the battery at time index $k$. The SoC, $z(k+1)$ is then calculated as:
\vspace{-3pt}\begin{equation}\label{cc-method}
 z(k+1) = z(k) - \left({\Delta}/{C}\right)\,I(k)
\vspace{-5pt}
\end{equation}
While SoC is sufficient to represent the battery state in single-cell battery packs, this is not necessarily true in multi-cell packs. In multi-cell packs, the battery SoC is decoupled from the energy stored in the battery pack because each cell in the pack may have different instantaneous SoC and even charge capacities. Due to this reason, we instead use the SoE to represent the battery state. 
The total available discharge energy of cell $i$ at time index $k$ is given as:
\vspace{-3pt}\begin{equation}\label{eq:energy-available}
    e_k^{(i)} = C \int_{z_{min}}^{z_k^{(i)}} OCV(\xi) \: \mathrm{d\xi}
\end{equation}
where $OCV$ is the Open-Circuit Voltage of the battery and $z_{min}$ is the minimum allowable SoC. This energy estimation exploits the non-linear chemical processes occurring in the battery that are manifested in the OCV-SoC relationship. We use the Neware BTS4000-5V battery cycler \cite{battery-cycler} to characterize the battery and derive the OCV-SoC relationship.
\begin{table}[tp]
\captionsetup{belowskip=18pt}
\caption{Energy consumption profile}
\vspace{-6pt}
\begin{center}
\begin{tabular}{|c|c|c|c|c|c|}
\hline
\textbf{Parameter} & \textbf{$I_{sl}$} & \textbf{$I_{sn}$} & \textbf{$I_{tx}$} & \textbf{$d_{sn}$} & \textbf{$d_{tx}$} \\
\hline
\textbf{Measurement} & 1.43 mA & 105 mA & 127 mA & 13s & 4.1s \\
\hline
\end{tabular}
\label{tab:energy-consump}
\end{center}
\vspace{-18pt}
\end{table}
\vspace{-12pt}
\subsection{Energy Consumption Profiling}
In order to predict the energy usage for selected sampling and transmission frequencies in the MPC framework, we must first determine the energy profile, i.e., the energy consumption for each individual task of the sensor node. We first describe the hardware configuration for which the energy profile is created. The microcontroller used is a LilyGo T-A7670E v1.2 wireless ESP32 development board \cite{esp32}. The sensors used are DS18B20 temperature sensor, PMS5003 particulate matter concentration sensor, and an MQ-135 gas sensor. The battery selected is a Molicel-INR18650-P26A \cite{molicel}.
\begin{figure}[b]
\vspace{-18pt}
\centering
\includegraphics[width=1\linewidth]{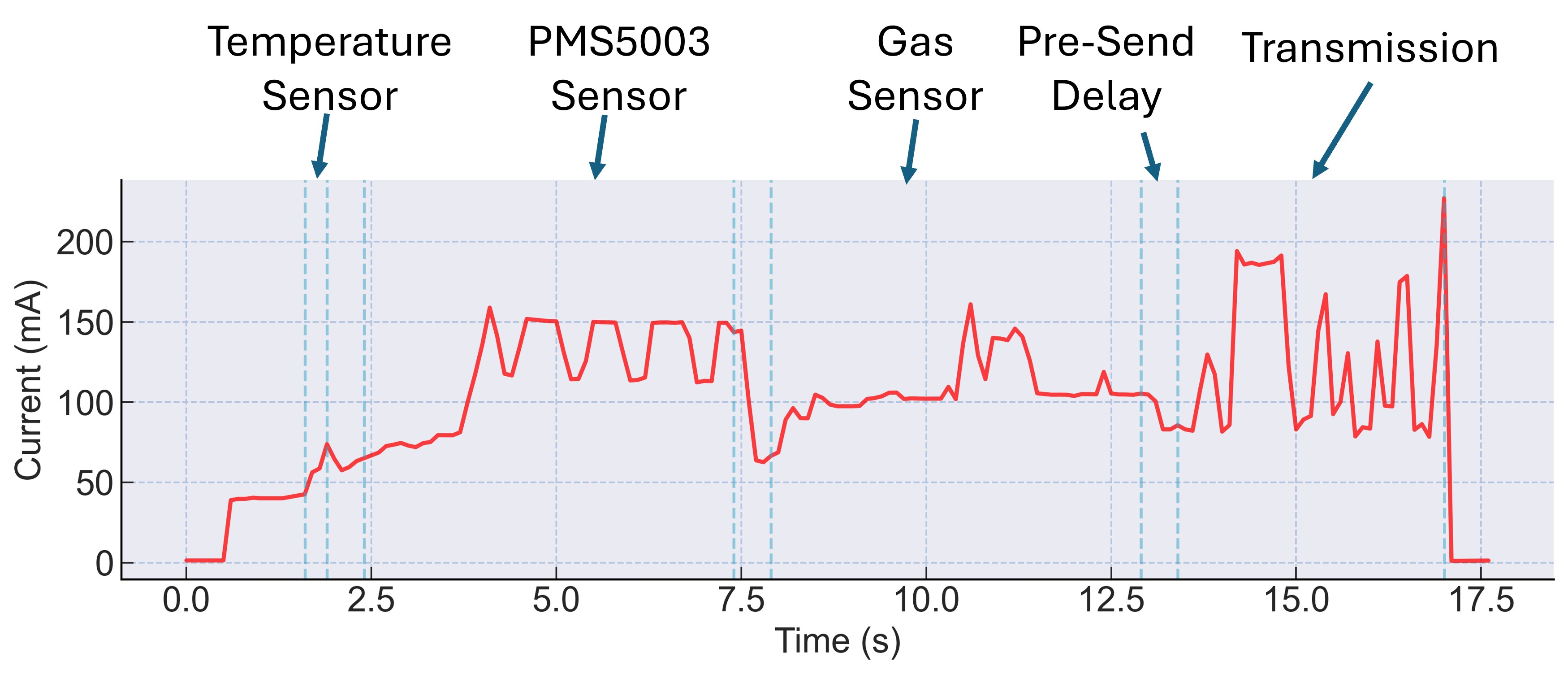}
\caption{The representative discharge profile. Note the short switching delay between sensors to stabilize before moving to the next sensing stage.}
\label{fig:discharge-profile}
\end{figure}

\begin{figure*}[t]
    \centering
    \begin{subfigure}{.49\textwidth}
        \includegraphics[scale=0.27]{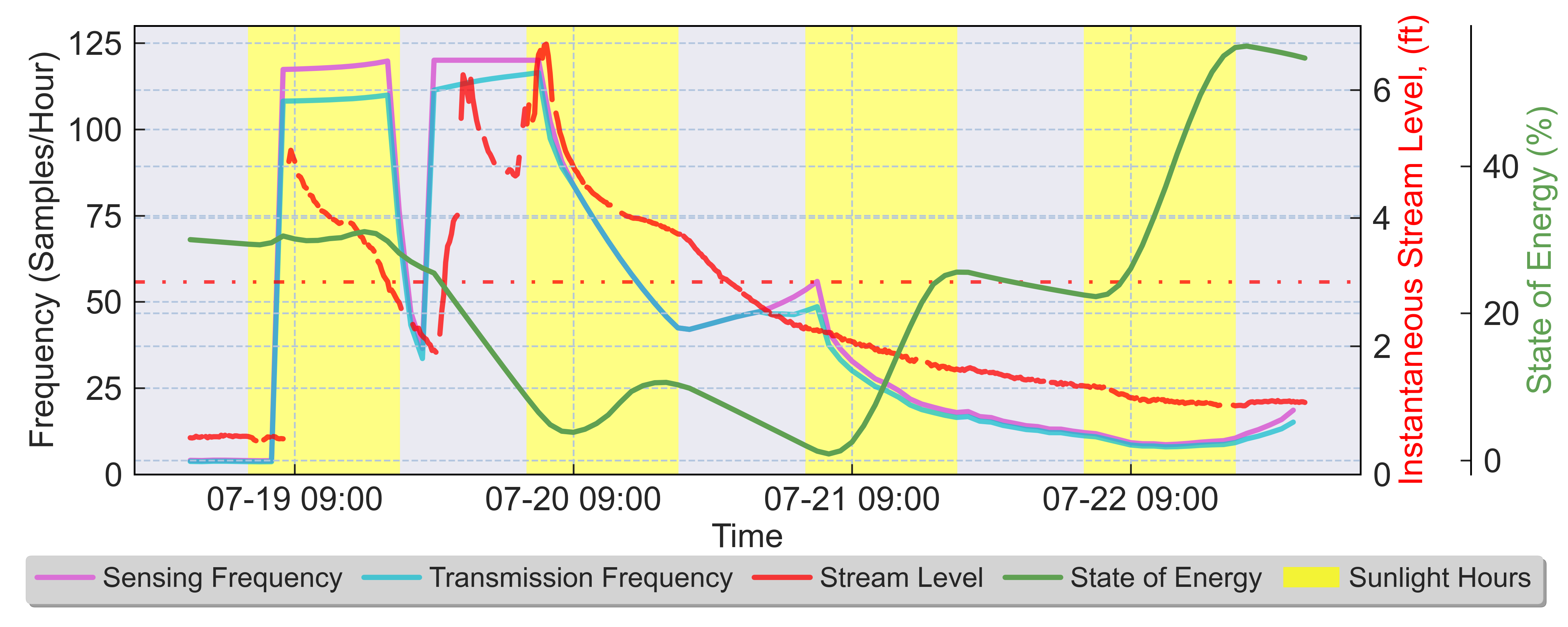}
        \caption{Initialized with $30\%$ SoE}
        \label{fig:control-action-init-soc-30}
    \end{subfigure}    
    \begin{subfigure}{.49\textwidth}
        \includegraphics[scale=0.27]{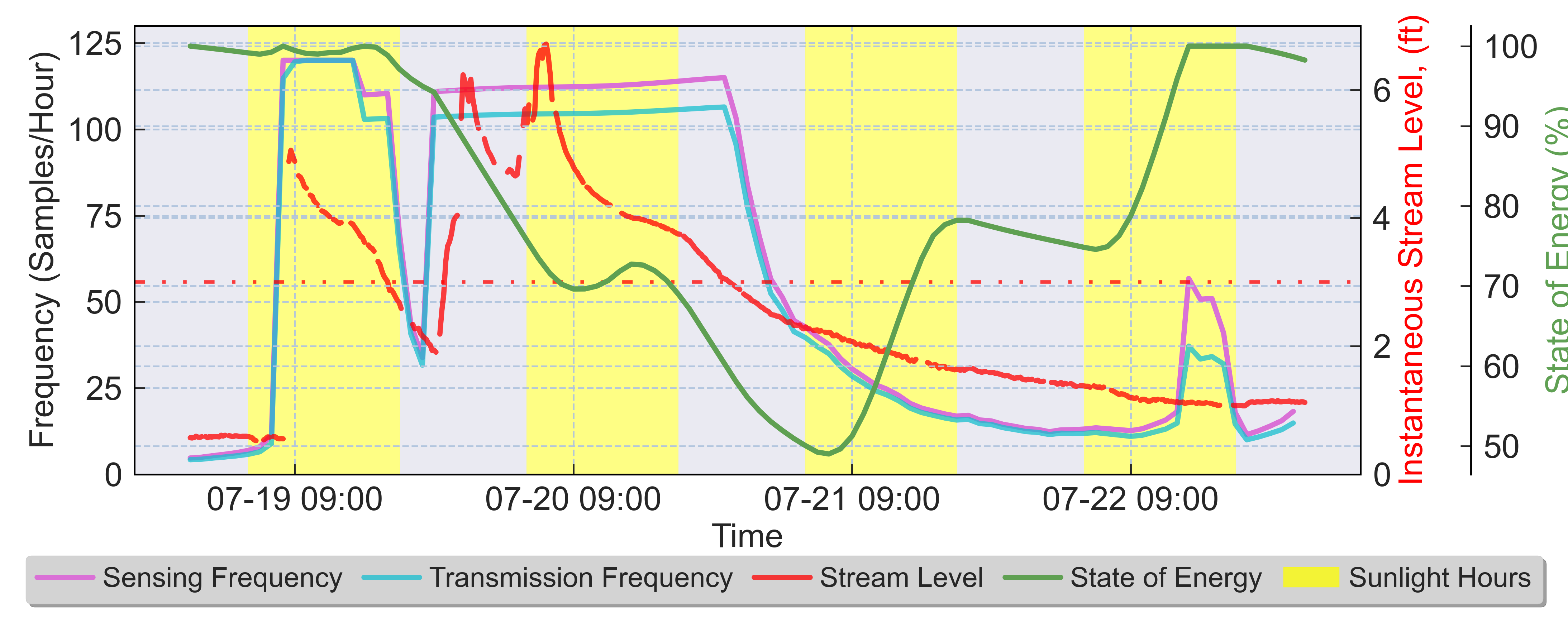}
        \caption{Initialized with $100\%$ SoE}
        \label{fig:control-action-init-soc-100}
    \end{subfigure}
    \caption{Control actions, stream levels, and SoE for different initial SoE values. Both VoI and SoE are equally weighted i.e. $w_v = w_e = 0.5$}
    \label{fig:control-actions-comparison}
    \vspace{-10pt}
\end{figure*}

Wireless sensor nodes operate in three phases: sleep, sense, and transmit. They predominantly remain in sleep, periodically waking up to acquire environmental data. Data transmission may occur immediately or be deferred. Energy consumption of a node varies by phase: sensing depends on the sensor type (e.g., PMS5003 and MQ-135 are energy-intensive), while transmission depends on protocol, signal strength, and data size.

Let $I_{\text{sl}}$, $I_{\text{sn}}$, $I_{\text{tx}}$ be the current drawn in the sleep, sense and transmit phases respectively. Also let the durations of the sense and transmit operations be $d_{\text{sn}}$, and $d_{\text{tx}}$. We can measure the current draw of each operating phase and its duration in an offline lab setting. It is assumed that this consumption profile remains consistent between cycles for a given set of hardware configuration \cite{tamkittikhun2017energy}. A representative sample of a full sleep-sense-transmit cycle is shown in Figure \ref{fig:discharge-profile}. 

The consumption profile obtained is given in Table \ref{tab:energy-consump}. To validate this profile, we run the given hardware at an arbitrarily chosen sensing and transmission frequency of 100 samples/hour.  Our experimental results show that at the given frequency and combination of hardware, the sensor node stops transmitting and sensing data after 49.8 hours. Our simulated results with the discharge profile in Table \ref{tab:energy-consump} estimates a shutdown time of 50.3 hours, an error of $0.994\%$.

\subsection{Mathematical Expression for State of Energy}
Using the Coulomb counting method and the energy consumption profile obtained previously, we now derive the state of energy, $0\leq S_e(f_s(k), f_t(k)) \leq 1$ as a function of the sampling and transmission frequencies $f_s(k)$ and $f_t(k)$. Let $\eta$ be the overall efficiency of the energy harvesting system, $P(k)$ be the  power harvested in $k$, and $V_{\text{nom}}$ be the nominal voltage of the battery. The remaining state of charge as a result of the decision set $(f_s(k), f_t(k))$ is given as
\begin{equation}
\label{eq:soc_f}
\begin{aligned}
\!\!\!\!z(k+1) = {} & z(k) - \frac{\Delta}{C}( f_s(k) I_{\text{sn}}d_{\text{sn}} + f_t(k) I_{\text{tx}}d_{\text{tx}}\\
& \hspace{-40pt}+ \!I_{\text{sl}}(1\!-\!f_s(k)d_{\text{sn}}\!-\!f_t(k)d_{\text{tx}}))\!+\! \frac{\eta}{C \!\times\! V_{nom}} \int_\Delta \!\!P(\tau)\: \mathrm{d\tau}
\end{aligned}
\end{equation}
SoE is then found as the ratio of the remaining energy capacity to the nominal energy capacity. It is given as
\begin{equation}
    S_e(f_s(k), f_t(k)) = \frac{\int_{z_{min}}^{z(k+1)} OCV(\xi) d\xi}{\int_{z_{min}}^{1} OCV(\xi) d\xi}
\end{equation}
where $z(k+1)$ is found from Equation \ref{eq:soc_f} and $OCV$ is found from the empirical OCV-SoC relationship obtained from the battery cycler as mentioned previously. The SoC is an affine and concave function.

\vspace{-1pt}\section{Results}\label{sec:results}

\begin{figure*}[t]
    \centering
    \begin{subfigure}{.49\textwidth}
        \includegraphics[scale=0.28]{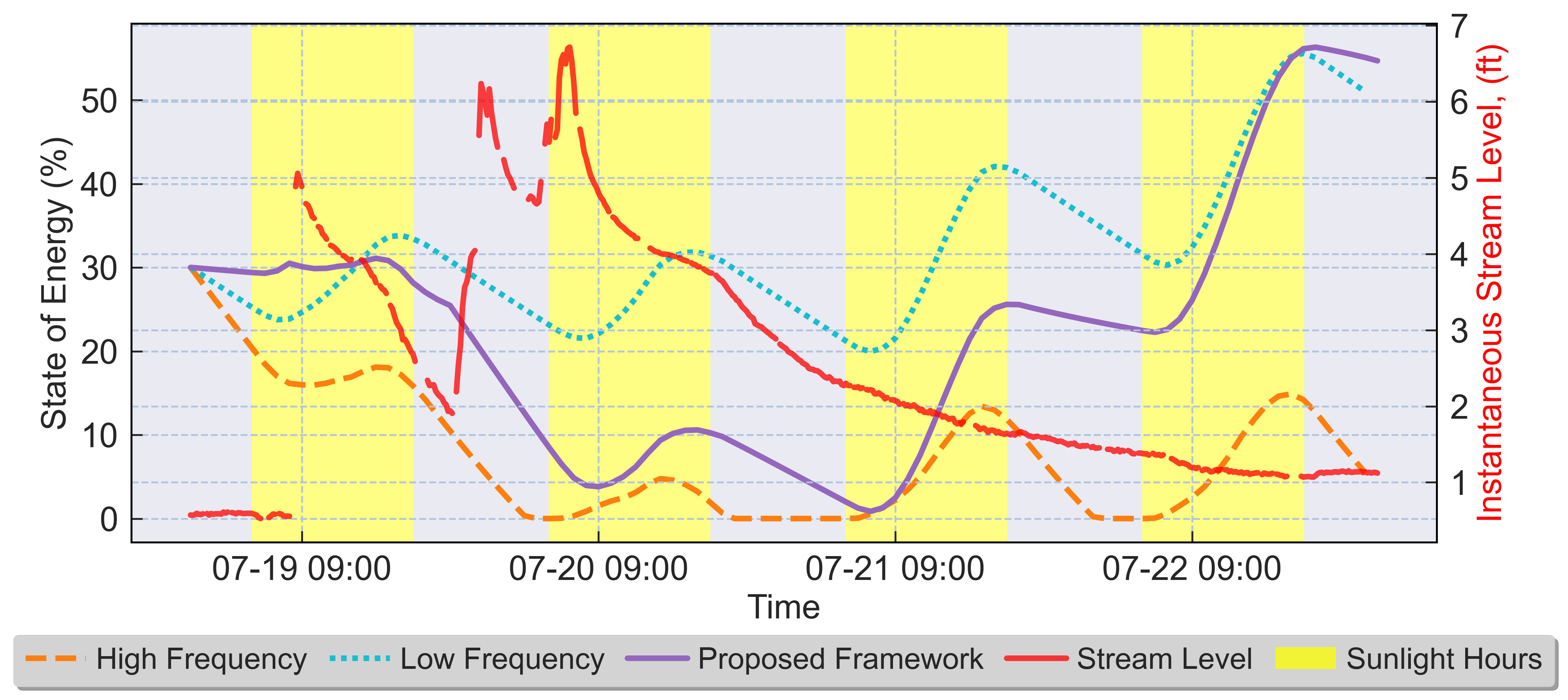}
        \caption{SoE of the node}
        \label{fig:baseline-soe}
    \end{subfigure}    
    \begin{subfigure}{.49\textwidth}
        \includegraphics[scale=0.28]{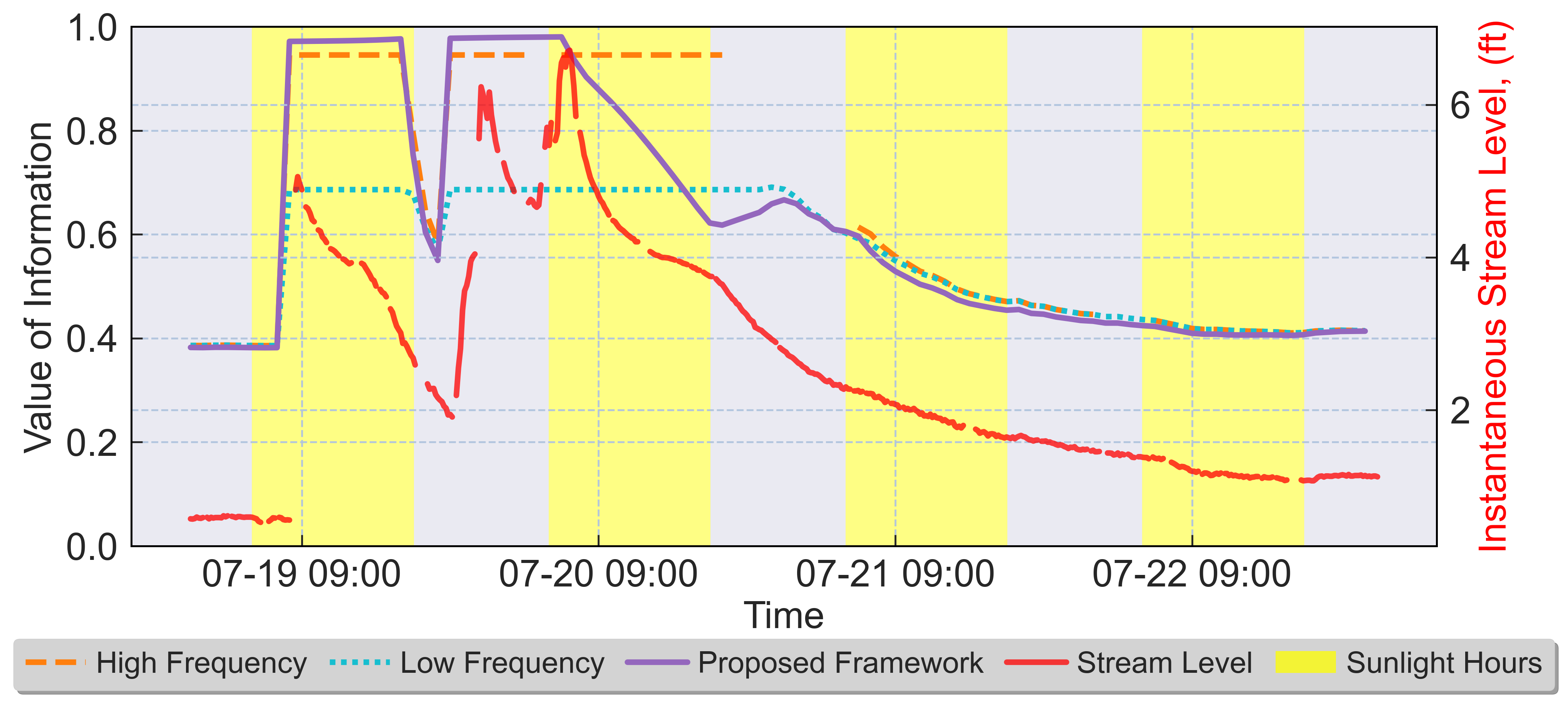}
        \caption{VoI of the data collected}
        \label{fig:baseline-voi}
    \end{subfigure}
    \caption{Baseline comparison of our proposed decision framework against static duty cycling algorithms when initialized with 30$\%$ SoE. Note that the gaps in the VoI of the high frequency plot correspond to regions with a depleted energy buffer as shown in the SoE plot.}
    \label{fig:baseline-comparison}
   \vspace{-10pt}
\end{figure*}
Here we present results to demonstrate the behavior of the MPC controller under different energy availability scenarios by hindcasting against data collected from a real-world sensor network \cite{manzoor2022development}. The data represents water level in an unstructured hill-torrent, and the time window considered includes a flash flood event. The sensor is installed on the Tarappi stream in an observation point in Lawa, Pakistan ($30\text{°} 22' 31.152'' N$ and $69\text{°} 20' 42.4176'' E$). In the end, we also compare the behavior of the MPC controller with fixed-frequency sampling and transmission at both high and low levels.
\vspace{-1pt}
\subsection{Simulation Setup}
\vspace{-1pt}
As mentioned previously, we simulate the behavior of the MPC controller against real-world stream level data. The parameter values are selected as follows: $\lambda_c = 1.4, x_c=3, \alpha_r = 0.018, \alpha_d = 0.025, D_0=0.5 ,\Delta=1, \zeta=0, \eta=0.05, C=2.75, \text{ and }z_{\text{min}}=0.015$. The prediction horizon $H_P$ is $11$. The maximum frequencies are $f_s^{max} = f_t^{max} = 120$. We use All Sky Surface Shortwave Downward Irradiance as a measure of solar irradiation received by assumed solar panels of size $10\text{ cm} \times 10\text{ cm}$ over the hindcasting window, obtained from the Prediction of Worldwide Energy Resources (NASA/POWER) project \cite{NASA_POWER}. The beliefs for the future values of the process (stream level) is obtained by assuming the maximum of the stream level over the previous window and for energy availability is obtained by assuming the solar radiation received over the prediction horizon $H_p$.
\vspace{-1pt}\subsection{Simulation Results}
\vspace{-1pt}Figures \ref{fig:control-action-init-soc-30} and \ref{fig:control-action-init-soc-100} showcase the controller's behavior under two different scenarios across the hindcasting window. First, in Figure \ref{fig:control-action-init-soc-30}, we see that despite the limited energy buffer at the start of the hindcasting window, the sensor chooses to operate at large sensing and transmission frequencies when a flash flood occurs (due to the high VoI). We also see that during the event of the second flash flood on 20th July, the sensor is first able to dynamically lower its frequencies (without entirely sacrificing the VoI objective) to conserve energy until the next harvesting opportunity, while also subsequently raising its frequency to fulfill the VoI objectives. We also see that due to the predictive model of energy harvest, the controller is able to preemptively increase its frequencies on the night of 20th July, despite having very little energy available. Next, in Figure \ref{fig:control-action-init-soc-100}, we see that the controller maintains a maximum sampling and transmission frequency throughout the duration of the flash flood event due to the abundant energy available in the energy buffer initially. A lower frequency to conserve energy is selected only when the threat proximity has subsided. Thus we see that the proposed decision framework is able to effectively balance the contrasting application and energy needs as envisioned.

Next, we compare the performance of the MPC controller against two static duty cycling algorithms operating at sensing and transmission frequencies of $(110, 90)$ and $(50,50)$ samples per hour, under tight energy constraints, shown in Fig. \ref{fig:baseline-comparison}. We see that we approach the high-frequency limit under high energy and high-threat conditions. Conversely, we approach the low-frequency limit under low energy and low-threat conditions. We see that overall, the MPC controller collects higher VoI data than the low frequency algorithm while simultaneously having an almost equal SoE at the end of the hindcasting window. While there are certain time windows where it collects less VoI than the high frequency node, the system is able to avoid any shutdowns due to a loss of energy. On the other hand, the high-frequency duty cycling algorithm faces long periods of inactivity due to insufficient energy.

\vspace{-1pt}\section{Conclusion}\label{sec:conclusion}
In this study we have focused on enhancing the autonomy in decision-making for IoT devices within both application and energy-related contexts. An MPC framework has been proposed to optimally navigate a sensor operating in a dynamic and unpredictable environment towards a finite, receding horizon. Our framework is able to conserve energy while simultaneously also improving the VoI of the data collected. While we have demonstrated the results for a specific application, our approach is generic enough to be applicable across a variety of domains. We aspire for this framework to drive further development in time-critical applications in environmental and industrial monitoring under tight energy constraints. \vspace{-1pt}
\bibliographystyle{ieeetr}
\bibliography{references}

@inproceedings{manzoor2022development,
  title={Development of Hydrometric Sensor Networks in Formerly Ungauged Watersheds: Lessons from Namal Valley, Mianwali},
  author={Manzoor, Talha and Arshad, Hassam and Nasir, Hasan Arshad and Sheraz, Muhammad and Khan, Malik Jahan},
  booktitle={IGARSS 2022-2022 IEEE International Geoscience and Remote Sensing Symposium},
  pages={7662--7665},
  year={2022},
  organization={IEEE}
}

@article{perera2013context,
  title={Context aware computing for the internet of things: A survey},
  author={Perera, Charith and Zaslavsky, Arkady and Christen, Peter and Georgakopoulos, Dimitrios},
  journal={IEEE communications surveys \& tutorials},
  volume={16},
  number={1},
  pages={414--454},
  year={2013},
  publisher={Ieee}
}

@inproceedings{giouroukis2020survey,
  title={A survey of adaptive sampling and filtering algorithms for the internet of things},
  author={Giouroukis, Dimitrios and Dadiani, Alexander and Traub, Jonas and Zeuch, Steffen and Markl, Volker},
  booktitle={Proceedings of the 14th ACM international conference on distributed and event-based systems},
  pages={27--38},
  year={2020}
}

@article{alfonso2021self,
  title={Self-adaptive architectures in IoT systems: a systematic literature review},
  author={Alfonso, Iv{\'a}n and Garc{\'e}s, Kelly and Castro, Harold and Cabot, Jordi},
  journal={Journal of Internet Services and Applications},
  volume={12},
  pages={1--28},
  year={2021},
  publisher={Springer}
}

@article{ali2017comprehensive,
  title={A comprehensive survey on real-time applications of WSN},
  author={Ali, Ahmad and Ming, Yu and Chakraborty, Sagnik and Iram, Saima},
  journal={Future internet},
  volume={9},
  number={4},
  pages={77},
  year={2017},
  publisher={MDPI}
}

@article{adu2018energy,
  title={Energy-harvesting wireless sensor networks (EH-WSNs) A review},
  author={Adu-Manu, Kofi Sarpong and Adam, Nadir and Tapparello, Cristiano and Ayatollahi, Hoda and Heinzelman, Wendi},
  journal={ACM Transactions on Sensor Networks (TOSN)},
  volume={14},
  number={2},
  pages={1--50},
  year={2018},
  publisher={ACM New York, NY, USA}
}

@article{kansal2007power,
  title={Power management in energy harvesting sensor networks},
  author={Kansal, Aman and Hsu, Jason and Zahedi, Sadaf and Srivastava, Mani B},
  journal={ACM Transactions on Embedded Computing Systems (TECS)},
  volume={6},
  number={4},
  pages={32--es},
  year={2007},
  publisher={ACM New York, NY, USA}
}

@inproceedings{vigorito2007adaptive,
  title={Adaptive control of duty cycling in energy-harvesting wireless sensor networks},
  author={Vigorito, Christopher M and Ganesan, Deepak and Barto, Andrew G},
  booktitle={2007 4th Annual IEEE communications society conference on sensor, mesh and ad hoc communications and networks},
  pages={21--30},
  year={2007},
  organization={IEEE}
}

@inproceedings{tamkittikhun2017energy,
  title={Energy consumption estimation for energy-aware, adaptive sensing applications},
  author={Tamkittikhun, Nattachart and Hussain, Amen and Kraemer, Frank Alexander},
  booktitle={Mobile, Secure, and Programmable Networking: Third International Conference, MSPN 2017, Paris, France, June 29-30, 2017, Revised Selected Papers 3},
  pages={222--235},
  year={2017},
  organization={Springer}
}

@article{donohoo2013context,
  title={Context-aware energy enhancements for smart mobile devices},
  author={Donohoo, Brad K and Ohlsen, Chris and Pasricha, Sudeep and Xiang, Yi and Anderson, Charles},
  journal={IEEE Transactions on Mobile Computing},
  volume={13},
  number={8},
  pages={1720--1732},
  year={2013},
  publisher={IEEE}
}

@article{ahituv1989assessing,
  title={Assessing the value of information: Problems and approaches},
  author={Ahituv, Niv},
  year={1989},
  journal={ICIS 1989 Proceedings},
  pages={45}
}

@article{kosta2020cost,
  title={The cost of delay in status updates and their value: Non-linear ageing},
  author={Kosta, Antzela and Pappas, Nikolaos and Ephremides, Anthony and Angelakis, Vangelis},
  journal={IEEE Transactions on Communications},
  volume={68},
  number={8},
  pages={4905--4918},
  year={2020},
  publisher={IEEE}
}

@inproceedings{padhy2006utility,
  title={A utility-based sensing and communication model for a glacial sensor network},
  author={Padhy, Paritosh and Dash, Rajdeep K and Martinez, Kirk and Jennings, Nicholas R},
  booktitle={Proceedings of the fifth international joint conference on Autonomous agents and multiagent systems},
  pages={1353--1360},
  year={2006}
}

@article{kho2009decentralized,
  title={Decentralized control of adaptive sampling in wireless sensor networks},
  author={Kho, Johnsen and Rogers, Alex and Jennings, Nicholas R},
  journal={ACM Transactions on Sensor Networks (TOSN)},
  volume={5},
  number={3},
  pages={1--35},
  year={2009},
  publisher={ACM New York, NY, USA}
}

@article{bisdikian2013quality,
  title={On the quality and value of information in sensor networks},
  author={Bisdikian, Chatschik and Kaplan, Lance M and Srivastava, Mani B},
  journal={ACM Transactions on Sensor Networks (TOSN)},
  volume={9},
  number={4},
  pages={1--26},
  year={2013},
  publisher={ACM New York, NY, USA}
}

@inproceedings{giordani2019framework,
  title={A framework to assess value of information in future vehicular networks},
  author={Giordani, Marco and Higuchi, Takamasa and Zanella, Andrea and Altintas, Onur and Zorzi, Michele},
  booktitle={Proceedings of the 1st acm mobihoc workshop on technologies, models, and protocols for cooperative connected cars},
  pages={31--36},
  year={2019}
}

@article{zhang2015value,
  title={Value of information aware opportunistic duty cycling in solar harvesting sensor networks},
  author={Zhang, Jianhui and Li, Zhi and Tang, Shaojie},
  journal={IEEE Transactions on Industrial Informatics},
  volume={12},
  number={1},
  pages={348--360},
  year={2015},
  publisher={IEEE}
}

@article{zhou2007floodnet,
  title={Floodnet: Coupling adaptive sampling with energy aware routing in a flood warning system},
  author={Zhou, Jing and De Roure, David},
  journal={Journal of Computer Science and Technology},
  volume={22},
  pages={121--130},
  year={2007},
  publisher={Springer}
}

@article{moser2009adaptive,
  title={Adaptive power management for environmentally powered systems},
  author={Moser, Clemens and Thiele, Lothar and Brunelli, Davide and Benini, Luca},
  journal={IEEE Transactions on Computers},
  volume={59},
  number={4},
  pages={478--491},
  year={2009},
  publisher={IEEE}
}

@inproceedings{braten2019adaptive,
  title={Adaptive, correlation-based training data selection for IoT device management},
  author={Braten, Anders Eivind and Kraemer, Frank Alexander and Palma, David},
  booktitle={2019 Sixth International Conference on Internet of Things: Systems, Management and Security (IOTSMS)},
  pages={169--176},
  year={2019},
  organization={IEEE}
}

@misc{esp32,
    author = "{LILYGO®}",
    title = "T-A7670E/G/SA R2",
    year = "2022",
    note = "[Online; accessed 4-March-2025]"
}

@misc{molicel,
    author = "{www.molicel.com}",
    title = "Explore by Product – Molicel,",
    url = "https://www.molicel.com/products-applications/explore-by-product/",
    note = "[Online; accessed 4-March-2025]"
}

@misc{NASA_POWER,
  author       = {{NASA Langley Research Center (LaRC) POWER Project}},
  year         = {2025},
  url          = {https://power.larc.nasa.gov/},
  note         = {Accessed: 28-February-2025}
}

@misc{battery-cycler,
    author = {Neware},
    title = {Neware BTS4000: The very adaptable and most popular battery cyclers/analyzers/testers in the whole battery world.},
    year = {2024},
    url = {https://newarebattery.com/neware-bts4000-series-customizable-output-power-range-and-functions/},
    note = {Accessed: 26-March-2025}
}

@article{kouvaritakis2016model,
  title={Model predictive control},
  author={Kouvaritakis, Basil and Cannon, Mark},
  journal={Switzerland: Springer International Publishing},
  volume={38},
  number={13-56},
  pages={7},
  year={2016},
  publisher={Springer}
}
\end{document}